\documentclass[aps,pra,twocolumn,floats,showpacs,floatfix]{revtex4}
\usepackage{epsfig}
\usepackage{color}
\usepackage{bm}
\usepackage{latexsym}

\begin{document}
\newcommand{\cc}{{\bf\Large C }}
\newcommand{\hide}[1]{}
\newcommand{\tbox}[1]{\mbox{\tiny #1}}
\newcommand{\half}{\mbox{\small $\frac{1}{2}$}}
\newcommand{\sinc}{\mbox{sinc}}
\newcommand{\const}{\mbox{const}}
\newcommand{\trc}{\mbox{trace}}
\newcommand{\intt}{\int\!\!\!\!\int }
\newcommand{\ointt}{\int\!\!\!\!\int\!\!\!\!\!\circ\ }
\newcommand{\eexp}{\mbox{e}^}
\newcommand{\EPS} {\mbox{\LARGE $\epsilon$}}
\newcommand{\ar}{\mathsf r}
\newcommand{\im}{\mbox{Im}}
\newcommand{\re}{\mbox{Re}}
\newcommand{\bmsf}[1]{\bm{\mathsf{#1}}}
\newcommand{\dd}[1]{\:\mbox{d}#1}
\newcommand{\abs}[1]{\left|#1\right|}
\newcommand{\bra}[1]{\left\langle #1\right|}
\newcommand{\ket}[1]{\left|#1\right\rangle }
\newcommand{\mbf}[1]{{\mathbf #1}}
\definecolor{red}{rgb}{1,0.0,0.0}

\title{Complexity in parametric Bose-Hubbard Hamiltonians and \\structural analysis of eigenstates}

\author{
Moritz Hiller,$^{1,2,3}$ Tsampikos Kottos,$^{1,2}$ and T. Geisel$^{2,3}$
}

\affiliation{
$^1$Department of Physics, Wesleyan University, Middletown, Connecticut 06459, USA \\
$^2$MPI for Dynamics and Self-Organization, Bunsenstra\ss e 10, D-37073 G\"ottingen, Germany\\
$^3$Fakult\"at f\"ur Physik, Universit\"at G\"ottingen, Friedrich-Hund-Platz 1, D-37077 G\"ottingen, Germany
}

\begin{abstract}
We consider a family of chaotic Bose-Hubbard Hamiltonians (BHH) parameterized by the coupling
strength $k$ between neighboring sites. As $k$ increases the eigenstates undergo changes, reflected 
in the structure of the Local Density of States. We analyze these changes, both numerically and 
analytically, using perturbative and semiclassical methods. Although our focus is on the quantum 
trimer, the presented methodology is applicable for the analysis of longer lattices as well.
\end{abstract}
\maketitle


Understanding the complicated behavior of quantum many-body systems of interacting Bosons
has been a major challenge for leading research groups over the last few years. In fact, the 
growing theoretical interest was further enhanced by recent experimental achievements. The 
most fascinating of these was the realization of Bose-Einstein condensation (BEC) of ultra-cold 
atoms in periodic optical lattices \cite{GMEHB02}, which allows for novel concrete 
applications of quantum mechanics such as atom interferometers and atom lasers. The flood 
of experimental realizations includes systems ranging from bond-excitations in molecules 
\cite{JGBSF05} to cantilever vibrations in micro-mechanical arrays \cite{SHSICC03} and 
Josephson arrays \cite{TMO00}. 

The simplest non-trivial model that describes interacting bosons on a lattice is the Bose-Hubbard 
Hamiltonian (BHH), which incorporates the competition between kinetic and interaction 
energy of the bosonic system. In a substantial part of the existing literature (see for 
example \cite{FPZ00}), the dynamics and spectral properties of BHH was investigated using 
a semiclassical picture. In contrast, quantum mechanical calculations of a BHH are often 
limited by severe computational memory restrictions. However, it is possible to treat small 
systems with two or three lattice sites. These studies are extremely relevant to bond 
excitations in small molecules \cite{SLE85}, few coupled Josephson junctions and BECs in 
optical traps with just a few wells \cite{SFGS97}. As to the experimental realization of 
the last case, microtrap technology \cite{R02} is probably the most promising approach for 
realizing these small systems. Remarkably, there is already an experimental realization 
\cite{AGFHCO05}, with promising applications.

In this context, the two-site system (dimer) has been analyzed thoroughly from both the 
semiclassical \cite{ELS85,TK88} and the purely quantum viewpoint \cite{FPZ00,BES90,KBK03}. 
Such investigations have revealed many interesting phenomena like the onset of $\pi$-phase 
oscillations, symmetry-breaking, and self-trapping of boson population, the latter being 
observed experimentally in \cite{AGFHCO05}. In fact, the dimer is integrable since there 
are two conserved quantities, the energy and the number of bosons. Nevertheless, the richness 
of the results provides a motivation to go beyond the dimer and consider new scenarios 
where even richer dynamics should be observed. The trimer opens new exciting opportunities,
in this respect, since the addition of a third site leads to (classically) chaotic behavior, 
thus paving the way to understand longer lattices. The trimer has been studied quite 
extensively in the semiclassical regime \cite{ETT95,FP02,CPC00}. Surprisingly enough, the 
quantum trimer \cite{FGS89,NHMM00} (not to mention longer lattices \cite{KB04}) is barely treated.
As a matter of fact, the majority of 
the quantum studies are focused on the statistical properties of levels \cite{FGS89}. 
However, spectral statistics is only the first step in understanding the behavior of a 
complex quantum system. Substantially more insight is gained through the study of eigenstates.


In this paper we consider the quantum trimer in the chaotic regime and study the structural 
changes that the eigenstates undergo as an experimentally controlled parameter $k$ 
(the coupling strength between neighboring sites) is varied. The object of our interest is 
the overlap of a given perturbed eigenstate $|n(k_0+\delta k)\rangle$ with the eigenstates 
$|m(k_0)\rangle$ of the unperturbed trimer
\begin{eqnarray} \label{e3}
\hspace*{-0.5cm}
P(n|m) = |\langle n(k_0+\delta k)|m(k_0)\rangle|^2 \,.
\end{eqnarray}
Alternatively, if regarded as a function of $n$ for a fixed $m$, the kernel $P(n|m)$ represents (up to some
trivial scaling) the Local Density of States (LDoS) \cite{HCGK06}. Its lineshape is fundamental for the
understanding of the associated dynamics since its Fourier transform is the so-called 
``survival probability amplitude". In our studies, we have identified three structural 
regimes of the $P(n|m)$, which are associated with two parametric scales defined as
\begin{equation}
\label{scales}
\delta k_{\rm qm}\propto {\tilde U}/N^{3/2} \quad {\rm and}\quad 
\delta k_{\rm prt} \propto {\tilde U}/N,
\end{equation}
where ${\tilde U}=NU$,  $N$ is the number of interacting bosons and $U$ is the on-site boson-boson 
interaction. For $\delta k<\delta k_{\rm qm}$ the perturbation mixes only neighboring levels: the 
main component of the kernel $P(n|m)$ remains unaffected while corrections are captured by standard 
textbook finite order perturbation theory.  For $\delta k_{\rm qm}<\delta k<\delta k_{\rm prt}$ 
a non-trivial structure appears, consisting of two distinct components: while the tails 
are still captured by perturbation theory, the central part is of non-perturbative nature and 
extends over an energy width $\Gamma\propto N^2\cdot\delta k^2/{\tilde U}$. For $\delta k>\delta 
k_{\rm prt}$, quantum mechanical perturbation theory fails totally. Instead, classical calculations 
can be used to predict the shape of $P(n|m)$. An overview of the parametric evolution of $P(n|m)$ 
is shown in Fig.~1.  


The trimeric Bose-Hubbard-Hamiltonian, which describes an interacting boson gas confined in a three
well lattice, is given in second quantization by:
\begin{equation}
\hat{H}=
0.5 U\sum_{i=1}^{3}{\hat n}_i ({\hat n}_i-1) -
k\sum_{i\neq j}\hat{b}_{i}^{\dagger}\hat{b}_{j};\quad \hbar=1.
\label{H3mode}
\end{equation}
In the BEC framework, $k=k_0+\delta k$, is the coupling strength between adjacent sites 
$i,j$, and can be controlled experimentally (in the context of optical lattices this can be 
achieved by adjusting the intensity of the laser beams that create the trimeric lattice), 
while $U=4 \pi\hbar^2a_sV_{\tbox{eff}}/m  $ describes the interaction between two atoms on a single site
($V_{\tbox{eff}}$ is the effective mode volume of each site, $m$ is the atomic mass, 
and $a_s$ is the $s$-wave scattering length of atoms). In the context of molecular physics \cite{JGBSF05,BES90} $k$ 
represents the electromagnetic and mechanical coupling between the bonds of adjacent molecules 
$i,j$, while $U$ represents the anharmonic softening of the bonds under extension. The 
operators ${\hat n}_i=\hat{b}_i^{\dagger}\hat{b}_i$ count the number of bosons at site~$i$; the 
annihilation and creation operators $\hat{b}_i$ and $\hat{b}_i^{\dagger}$ obey the canonical commutation 
relations $[\hat{b}_i,\hat{b}_j^{\dagger}]=\delta_{i,j}$. Hamiltonian (\ref{H3mode}) 
has two constants of motion, namely the energy $E$ and the total number of particles $N=
\sum_{i=1}^3n_{i}$. Having $N=const.$ implies a finite Hilbert-space of dimension ${\cal 
N}=(N+2)(N+1)/2$ \cite{FGS89,BES90}. An additional 3-fold permutation symmetry allows us
to reduce further the dimensionality of our space \cite{FGS89}. 


When $N\gg1$ one can adopt a semiclassical point of view for Hamiltonian (\ref{H3mode}).
Formally, this can be seen if we define rescaled creation and annihilation operators $\hat{c}_i
=1/\sqrt{N}\hat{b}_i$. The corresponding commutators $[\hat{c}_i,\hat{c}_j^{\dagger}]=\delta_{ij} /N$ vanish
for $N\gg1$ and therefore one can treat the rescaled operators as c-numbers. The classical 
Hamiltonian $H$ is obtained using the Heisenberg relations $\hat{c}_i\rightarrow \sqrt{I_i}\exp^{i
\varphi_{i}}$ where $\varphi_i$ is an angle and $I_i$ is the associated action. We then get
\begin{equation}
{\tilde {\cal H}} = \frac{\cal H}{N{\tilde U}} =\frac{1}{2}\sum_{i=1}^{3}I_{i}^{2}-
\lambda\sum_{i\neq j}{\sqrt {I_{i}I_{j}}}\exp^{i(\varphi_j-\varphi_i)}\, .
\label{eq:H-DNLS}
\end{equation}
The dynamics is obtained from (\ref{eq:H-DNLS}) 
using the canonical equations $dI_i/d{\tilde t} = -\partial {\tilde {\cal H}}/\partial \varphi_i$ 
and $d\varphi_i/d{\tilde t} = \partial {\tilde {\cal H}}/\partial I_i$. Here ${\tilde t}={\tilde U} 
\cdot t$ is the rescaled time. The dimensionless ratio $\lambda\equiv k/{\tilde U}$ 
\cite{SLE85,TK88,FGS89,FP02,NHMM00} determines the dynamics of the classical Hamiltonian
(\ref{eq:H-DNLS}). For $\lambda\rightarrow 0$ the interaction term dominates and the system 
behaves as a set of uncoupled sites while for $\lambda\rightarrow \infty$ the kinetic term 
is the dominant one. In both limits the motion is integrable. We consider intermediate values 
of $\lambda$ where the classical dynamics is chaotic. 
The semiclassical limit is approached by keeping $\lambda_0$ and ${\tilde U}$ 
constant while $N\rightarrow\infty$. This is crucial in order to keep the underlying classical 
motion unaffected.

We assume that both ${\tilde {\cal H}}_0 = {\tilde {\cal H}}(\lambda_0)$ and ${\tilde {\cal 
H}} = {\tilde {\cal H}}(\lambda)$ generate classically chaotic dynamics of similar nature. 
This is equivalent with the requirement that $\delta \lambda \equiv (\lambda{-}\lambda_0)$ is 
{\em classically small}, i.e., $\delta \lambda \ll \delta \lambda_{ \rm cl}$. An important 
classical observable is the generalized force ${\tilde {\cal F}}({\tilde t}) 
\equiv -(\partial {\tilde {\cal H}}/\partial \lambda) = \sum_{i\neq j} {\sqrt {I_{i}I_{j}}}
\exp^{i (\varphi_j-\varphi_i)}$. Due to the chaotic dynamics the power spectrum 
$\tilde{C}({\tilde \omega})$ has a 
finite support ${\tilde \Omega}_{\rm cl}=2\pi/\tilde {t}_{\rm erg}$ with $\tilde {t}_{\rm 
erg}$ being the ergodic time of the system described by Eq.~(\ref{eq:H-DNLS}). 
One can use ${\tilde C}({\tilde \omega})$ as an operative way to evaluate $\delta \lambda_{
\rm cl}$. The latter is the maximum perturbation which keeps ${\tilde C}({\tilde \omega})$ unaffected. 
We note that in the experiment, the relevant parameter that is tuned is the coupling strength 
$\delta k_{\rm cl}={\tilde U} \cdot \delta \lambda_{\rm cl}$. We have found that for 
$0.04\leq \lambda\leq 0.2$ and in an energy interval ${\tilde H}\approx 0.26\pm 0.02$ the 
motion is predominantly chaotic. In this regime, and for ${\tilde U}=280$ we get $\delta 
k_{\rm cl}=20$ while ${\tilde \Omega}_{\rm cl}\approx 1$ (see Fig.~2a).


Quantum mechanically, we work in the eigenbasis of the Hamiltonian ${\hat H}_0$. In this basis
${\hat H}_0$ becomes diagonal i.e. $\mbf{E}_0=E_m^{(0)}\delta_{mn}$ where $\{E_m^{(0)}\}$ are the 
ordered eigenvalues. Their mean level spacing $\Delta \approx 1.5 {\tilde U}/N$ can be estimated 
using ${\cal N}\propto N^2$ together with Eq.~(\ref{eq:H-DNLS}). The perturbed Hamiltonian 
${\hat H}$ is written as
\begin{eqnarray} \label{e1}
\mbf{H} \ \  = \ \ \mbf{E}_0 \ - \ \delta k \ \mbf{B}
\end{eqnarray}
We mark that although the perturbation strength $\delta k$ is assumed to be classically small
($\delta k\leq \delta k_{\rm cl}$), quantum-mechanically it can be very {\em large}, i.e., it 
can mix many levels (see the avoided crossings appearing in Fig.~1 for large values of $\delta 
k$). From exact diagonalization we find that $\mbf{B}$ is a {\em banded matrix}. Its bandprofile
can be determined using a semiclassical recipe 
$\langle|\rm{B}_{nm}|^2\rangle \approx (N^2/{\tilde U})\Delta \cdot \tilde{C}(\omega_{nm})/2\pi$
\cite{FP86}. The bandwidth
is $\Delta_b={\tilde \Omega}_{\rm cl} \cdot {\tilde U}$. It is common to define $b\equiv\Delta_b/
\Delta \approx 0.6 N$. The banded matrix $\mbf{B}$ and the band profile are illustrated in Fig.~2a.


A fixed assumption of this work is that $\delta k\ll \delta k_{\rm cl}$. If we require the 
perturbation to be also {\em quantum mechanically small} $\delta k\leq \delta k_{\rm qm}$, 
then we can apply standard first order perturbation theory (FOPT) (see Fig.~2b). In this case 
$P_{\tbox{FOPT}} (n|m)\approx 1$ for $n=m$, while
$P_{\tbox{FOPT}}(n|m) = {\delta k^2 \ \langle|\rm{B}_{nm}|^2\rangle}/{(E_n{-}E_m)^2}.$
The perturbation strength $\delta k_{\rm qm}$ is given by the condition that only neighboring 
levels are mixed, yielding $\delta k_{\rm qm}=\Delta /\sqrt{\langle |\rm{B}_{nm}|^2}\rangle$. Substituting
the expressions for $\Delta$ and $\langle |\rm{B}_{nm}|^2\rangle$ we obtain the results reported in
Eq.~(\ref{scales}). In Fig.~3 we report our numerical results for $\delta k_{\rm qm}$ \cite{note2}. 
A nice agreement with Eq.~(\ref{scales}) is observed.

For stronger perturbations $\delta k_{\rm qm} < \delta k$ one has to employ perturbation 
theory to infinite order. Until now, the only formal results regarding Hamiltonians 
of the type (\ref{e1}), have been derived by Wigner \cite{W55}. 
He assumed that $\rm{B}_{nm}$ is a {\it Banded Random Matrix} with a flat band profile and
found that $P(n|m)$ is a Lorentzian
\begin{equation}
\label{pert}
P_{\rm prt}(n|m) =
{\delta k^2 |\mbf{B}_{nm}|^2 \over \Gamma^2 + (E_n{-}E_m)^2},\quad 
\Gamma=  {\langle|\mbf{B}_{nm}|^2\rangle \over \Delta}\delta k^2.
\end{equation}
For $\Gamma\leq \Delta$ Eq.~(\ref{pert}) reduces to $P_{\tbox{FOPT}}$ 
while for $\Delta< \Gamma <\Delta_b$ the kernel $P(n|m)$ contains two 
distinct components: a central region $|E_n-E_m|\leq \Gamma$ where the mixing of levels 
is non-perturbative and a tail region $\Gamma<|E_n-E_m|<\Delta_b$ which can still
be captured by perturbation theory. The condition $\Gamma\sim \Delta_b$ determines
$\delta k_{\rm prt}$ above which the non-perturbative core extends all over the bandwidth.
Therefore (\ref{pert}) applies as long as $\delta k <\delta k_{\rm prt}\propto \delta 
k_{\rm qm} \cdot b^{0.5}$. 
In the case of non-interacting systems with chaotic classical limit recent studies \cite{CH00,CK01}
indicated that the above scenario, based on random matrix modeling, leads to a fairly
good description of the kernel. Specifically, $P(n|m)$ was found to exhibit a core-tail
structure where the width of the core scaled as $\Gamma\sim \delta k^{\alpha}$ 
with $1<\alpha\le 2$ \cite{CH00} in contrast to Eq.~(\ref{pert}). In these studies the 
parameter $\Gamma$ was determined (for a given $\delta k$) by imposing normalization of 
$P_{\rm prt}(n|m)$. We note that a strict Lorentzian is an idealization of the random 
matrix modeling.

Does our BHH model follow the same scenario or will the interactions affect the shape of $P(n|m)$?
Already from Fig.~1 we see that as $\delta k>\delta k_{\rm qm}$, a non-perturbative core starts 
to evolve and eventually spills over the whole bandwidth $\Delta_b$. A more detailed comparison 
(see Fig.~2) between $P(n|m)$ and Eq.~(\ref{pert}) shows an excellent agreement. In the inset 
of Fig.~3 we report the measured width $\Gamma$ as a function of $\delta k$. 
Our numerical data are in agreement with a power law behavior $\Gamma\propto \delta k^{\alpha}$ 
with $\alpha = 2\pm 0.01$. We then investigated the scaling behavior of $\delta k_{\rm prt}$ 
\cite{note2}. Our numerical data are reported
in Fig.~3 and are in agreement with the results stated in Eq.~(\ref{scales}). The latter
equation can now be understood in the light of Wigner's arguments since $\delta k_{\rm prt}
\propto \delta k_{\rm qm} \times b^{0.5}$ and $b\sim N$.


For $\delta k>\delta k_{\rm prt}$ the core spills over the bandwidth and therefore 
perturbation theory, even to infinite order, is inapplicable for evaluating $P(n|m)$. 
In such cases one has to rely on completely non-perturbative methods \cite{CH00,CK01}. 
For the Wigner model, it was found that $P(n|m) = 1/(2\pi\Delta) \sqrt{4-((E_n-E_m)/\Delta)^2}$
\cite{W55} while in systems that have a semiclassical 
limit the overlap kernel becomes $P(n|m)= {\rm tr}\left( \rho_n \rho_m\right)$. Here 
$\rho_m(\{I_i\},\{\varphi_i\})$ and $\rho_n(\{I_i\},\{\varphi_i\})$ are the Wigner functions 
that correspond to the eigenstates $|m(k_0)\rangle$ and $|n(k)\rangle$ respectively.
The trace stands for $d\{I_i\}d\{\varphi_i\}/(2\pi\hbar)^d$ integration. In the classical 
limit $\rho$ can be approximated by the corresponding
micro-canonical distribution $\rho \propto \delta(E{-}{\cal H}(\{I_i\},\{\varphi_i\}))$. 
The latter can be evaluated by projecting the dynamics generated by ${\cal H}_0(\{I_i\},\{\varphi_i\})=E_0$
onto the Hamiltonian ${\cal H}(\{I_i\}, \{\varphi_i\})=E(t)$. In Fig.~4a we plot the resulting 
$E(t)$ as a function of time. The classical distribution $P(n|m)$ is constructed (Fig.~4b) from 
$E(t)$, averaged over a sufficiently long time. 

In conclusion, we have analyzed the structural changes which the eigenstates of a trimeric BHH undergo
as the coupling strength $\delta k$ between the neighboring sites is varied. We have found that for 
$\delta k < \delta k_{\rm prt}$ perturbation theory (to infinite order if $\delta k>\delta k_{\rm qm}$)
is applicable. In this case, the power spectrum of the generalized force $\tilde{C}({\tilde \omega})$ is an
important ingredient for the theory. It is directly experimentally measurable \cite{Z03} because the 
momentum distribution of atoms in a lattice is $\sim \sum_j \exp(ikj) X_j$ where $k$ is the atomic momentum
and $X_j=\langle [{\hat b}_{j+l}^{\dagger}{\hat b}_l+{\rm h.c}]\rangle$ is the one-particle density matrix.
In the opposite limit $\delta k>\delta k_{\rm prt}$ one can apply semiclassical (non-perturbative) 
considerations. 


It is our pleasure to acknowledge very fruitful discussions with 
D. Cohen and G. Kalosakas.
\begin{figure}
\includegraphics[width=\columnwidth,keepaspectratio,clip]{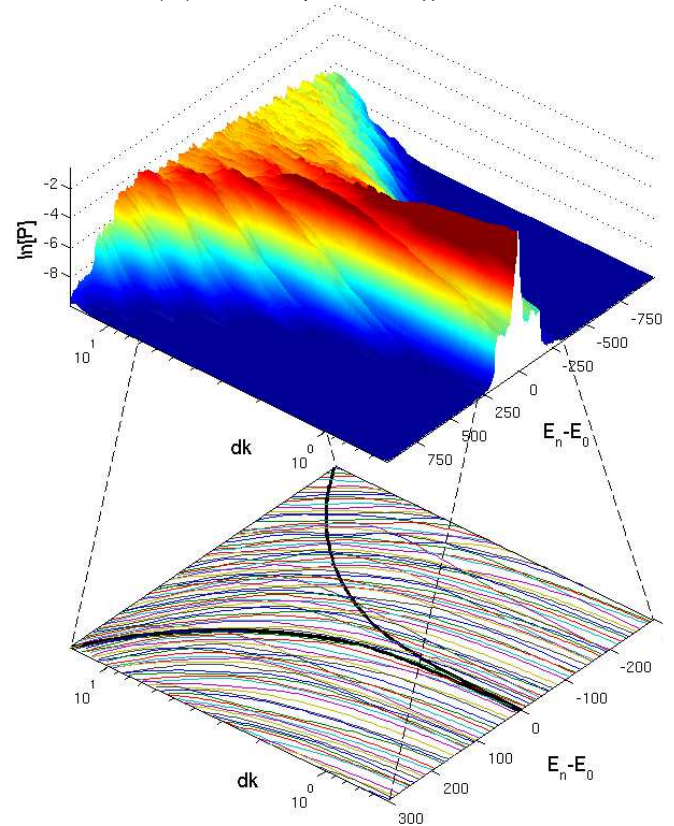} 
\caption{\label{fig1}
(Color online) The kernel $P(n|m)$ of the BHH plotted as a function of the perturbed energies $E_n$ (LDoS representation) 
and for various perturbation strengths $\delta k>\delta k_{\rm qm}$. In the lower plane we report the parametric
evolution of the energy levels within the bandwidth $\Delta_b\approx 230$. The energy width $\Gamma$ is shown
(bold line) as a function of $\delta k$. The averaged shape of eigenfunctions is given by the same kernel
$P(n|m)$ and is obtained by just inverting the energy axis. Here, $N=70$, and $\lambda_0=0.053$.      
} 
\end{figure}

\begin{figure}
\includegraphics[width=\columnwidth,keepaspectratio,clip]{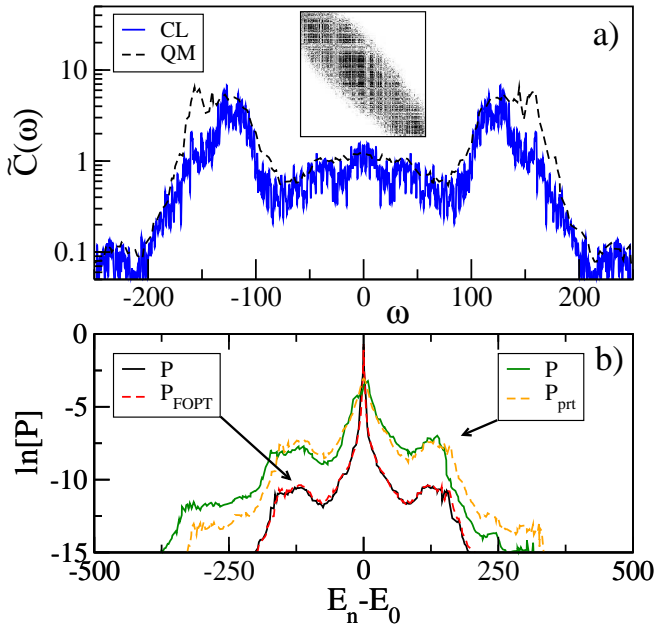} 
\caption{\label{fig2}
(Color online){\em Upper panel}: Comparison of the quantum band profile 
$\langle|\rm{B}_{nm}|^2\rangle ({2\pi\tilde U}/N^2 \Delta)$
with the classical power spectrum $\tilde{C}(\omega)$. 
The number of particles is $N=230$ and $\lambda_0=0.053$. The inset shows a snapshot of the matrix {\bf B}.   
{\em Lower panel}: The kernel $P(n|m)$ of the BHH (see Eq.(\ref{e3})) for $\delta k=0.05$
and for $\delta k=0.3$ compared to the corresponding theoretical expressions $P_{\tbox{FOPT}}$ 
and $P_{\rm prt}$. Here $\delta k_{\rm qm}=0.09$ and $\delta k_{\rm prt}=1.02$.
} 
\end{figure}

\begin{figure}
\includegraphics[width=\columnwidth,keepaspectratio,clip]{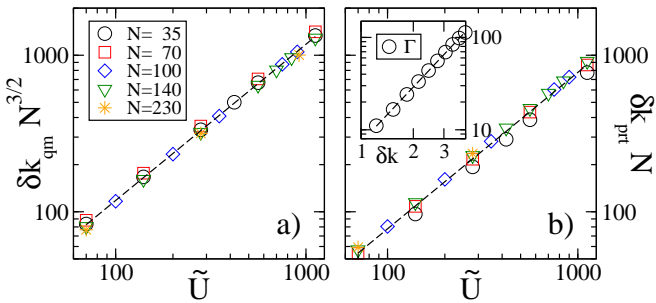} 
\caption{\label{fig3}
(Color online) The parameters (a) $\delta k_{\rm qm}$ and (b) $\delta k_{\rm prt}$ for various $\tilde{U},N$ 
and for $\lambda_0=0.053$. A nice scaling in accordance with Eq.(\ref{scales}) is observed. 
In the inset of (b) we report the scaling of the energy width $\Gamma$ with $\delta k$. 
The dashed line has slope $2$.
} 
\end{figure}

\begin{figure}
\includegraphics[width=\columnwidth,keepaspectratio,clip]{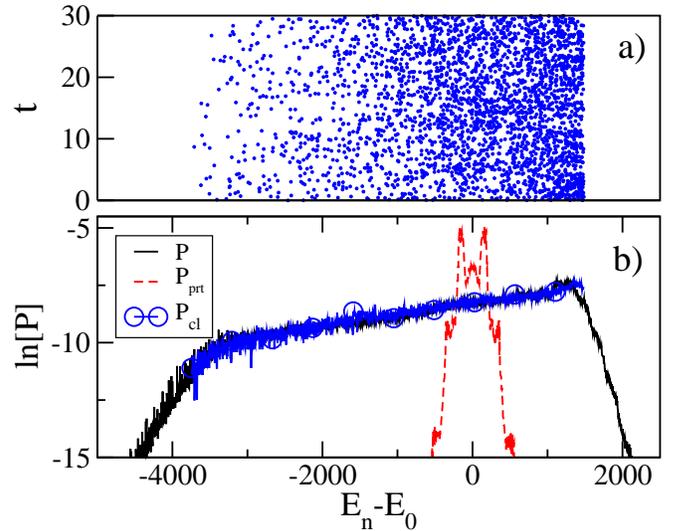} 
\caption{\label{fig4}
(Color online) The kernel $P(n|m)$ (LDoS representation) for the BHH in the non-perturbative regime ($\delta k=10$)
for $N=230$ and $\lambda_0=0.053$.~In the upper panel we plot a time series $E(t)$ which leads to the
classical profile $P_{\rm cl}(E)$ (see text for details).
} 
\end{figure}



\begin{thebibliography}{99}

\bibitem{GMEHB02} M. Greiner {\em et al.}, Nature {\bf 415}, 39 (2002); 
D. Jaksch {\em et al.}, Phys. Rev. Lett. {\bf 81}, 3108 (1998).

\bibitem{JGBSF05} M. Joyeux {\em et al.}, Adv. Chem. Phys.{\bf 130}, 267 (2005);
K. Lehmann, {\it et al.}, Annu. Rev. Phys. Chem. {\bf 45}, 241 (1994).

\bibitem{SHSICC03} M. Sato {\em et al.}, Phys. Rev. Lett. {\bf 90}, 044102 (2003).

\bibitem{TMO00} E. Tr\'ias, J. J. Mazo, and T. P. Orlando, Phys. Rev. Lett. {\bf 84}, 741 (2000);
P. Binder {\em et al.}, ibid. {\bf 84}, 745 (2000).

\bibitem{FPZ00} R. Franzosi, V. Penna and R. Zecchina, Int. J. Mod. Phys. B {\bf 14}, 943 (2000). 

\bibitem{SLE85} A. C. Scott, P. S. Lomdahl, J. C. Eilbeck, Chem. Phys. Lett. {\bf 113}, 29 (1985);
A. C. Scott, J. C. Eilbeck, Chem. Phys. Lett. {\bf 132}, 23 (1986); A. C. Scott, L. Bernstein,
J. C. Eilbeck, J. Biol. Phys. {\bf 17}, 1 (1989).

\bibitem{SFGS97} A. Smerzi {\em et al.}, Phys. Rev. Lett. {\bf 79},4950 (1997);
 S. Raghavan {\em et al.}, Phys. Rev. A {\bf 59}, 620 (1999).

\bibitem{R02} J. Reichel, Appl. Phys. B-Lasers and Optics {\bf 74}, 469 (2002).

\bibitem{AGFHCO05} M. Albiez {\em et al.}, Phys. Rev. Lett. {\bf 95}, 010402 (2005).

\bibitem{ELS85} J. C. Eilbeck, P. S. Lomdahl, A. C. Scott, Physica D {\bf 16}, 318 (1985);
E. Wright {\em et al.}, Physica D {\bf 69}, 18 (1993).

\bibitem{TK88} G. P. Tsironis, V. M. Kenkre, Phys. Lett. A {\bf 127}, 209 (1988); V. M. Kenkre,
G. P. Tsironis, Phys. Rev. B {\bf 35}, 1473 (1987); V. M. Kenkre, D. K. Campbell, Phys. Rev. B
{\bf 34}, R4959 (1986).

\bibitem{BES90} L. Bernstein, J. Eilbeck, A. Scott, Nonlinearity {\bf 3}, 293 (1990);
S. Aubry {\rm et al.}, Phys. Rev. Lett. {\bf 76}, 1607 (1996).

\bibitem{KBK03} G. Kalosakas, A. R. Bishop, V. M. Kenkre, Phys. Rev. A {\bf 68}, 023602 (2003);
G. Kalosakas, A. R. Bishop, Phys. Rev. A {\bf 65}, 043616 (2002); 
G. J. Milburn {\em et al.}, Phys. Rev. A {\bf 55}, 4318 (1997). 

\bibitem{ETT95}  J. C. Eilbeck, G. P. Tsironis, S. K. Turitsyn, Physica Scripta {\bf 52}, 
386 (1995); D. Hennig {\em et al.}, Phys. Rev. E {\bf 51}, 2870 (1995).

\bibitem{FP02} R. Franzosi, V. Penna, Phys. Rev. A {\bf 65}, 013601 (2002); 
R. Franzosi, V. Penna, Phys. Rev. E {\bf 67}, 046227 (2003). 

\bibitem{CPC00} L. Casetti, M. Pettini, E. G. D. Cohen, Phys. Rep. {\bf 337}, 237 (2000).

\bibitem{FGS89} S. de Filippo, M. Fusco Girard, M. Salerno, Nonlinearity {\bf 2}, 477 (1989);
A. Chefles, J. Phys. A {\bf 29}, 4515 (1996); S. Flach, V. Fleurov, J. Phys.: Cond. Matt. 
{\bf 9}, 7039 (1997).

\bibitem{NHMM00} K. Nemoto {\em et al.}, Phys. Rev. A {\bf 63}, 013604 (2000).

\bibitem{KB04} A.R. Kolovsky, A. Buchleitner, Europhys. Lett. {\bf 68}, 632 (2004).

\bibitem{HCGK06} M. Hiller {\em et al.}, Ann. Phys. {\bf 312}, 1025 (2006).

\bibitem{FP86} M. Feingold and A. Peres, Phys. Rev. A {\bf 34} 591, (1986).

\bibitem{note2} In our numerical analysis we have defined, $\delta k_{\rm qm}$ as the perturbation 
strength for which $50\%$ of the probability remains at the original site, while $\delta k_{\rm prt}$
is the $\delta k$ for which the variance of the perturbative expression (\ref{pert}) becomes less 
than $80\%$ of the exact variance.

\bibitem{W55} E. Wigner, Ann. Math. {\bf 62}, 548 (1955); {\bf 65}, 203 (1957); Y. V. 
Fyodorov, {\it et al.}, Phys. Rev. Lett. {\bf 76}, 1603 (1996).

\bibitem{CH00} 
D. Cohen and E. J. Heller, Phys. Rev. Lett. {\bf 84}, 2841 (2000); A. Barnett, D. Cohen and 
E.J. Heller, ibid. {\bf 85}, 1412 (2000); D. Cohen, Ann. Phys. {\bf 283}, 175-231 (2000).

\bibitem{CK01} 
J. A. M\'endez-Berm\'udez, T. Kottos and D. Cohen, Phys. Rev. E {\bf 72}, 027201 (2005);
D. Cohen and T. Kottos, Phys. Rev. E {\bf 63}, 036203 (2001). 

\bibitem{Z03} W. Zwerger, J. Opt. B: Quantum Semiclass. Opt. {\bf 5}, S9 (2003).
\end{thebibliography}
\end{document}